\documentclass[prl,reprint,twocolumn,showpacs,superscriptaddress]{revtex4-1}

\usepackage{bm,mathrsfs}
\usepackage{graphicx}
\usepackage{epsfig}
\usepackage{amsmath,bbm}
\usepackage{amsfonts,amssymb}
\usepackage{times}
\usepackage{dsfont}
\usepackage{enumitem}  
\usepackage{comment}
\usepackage[colorlinks,linkcolor=blue,citecolor=blue]{hyperref}
\newcommand{\eqn}[1]{
\begin{eqnarray}
	#1
\end{eqnarray}
}

\begin{document}

\title{Full-Counting Many-Particle Dynamics: Nonlocal and Chiral Propagation of Correlations}

\author{Yuto Ashida}
%\email{ashida@cat.phys.s.u-tokyo.ac.jp}
\affiliation{Department of Physics, University of Tokyo, 7-3-1 Hongo, Bunkyo-ku, Tokyo
113-0033, Japan}
\author{Masahito Ueda}
%\email{ueda@phys.s.u-tokyo.ac.jp}
\affiliation{Department of Physics, University of Tokyo, 7-3-1 Hongo, Bunkyo-ku, Tokyo
113-0033, Japan}
\affiliation{RIKEN Center for Emergent Matter Science (CEMS), Wako, Saitama 351-0198, Japan
}

\begin{abstract}
The ability to measure single quanta has allowed complete characterization of small quantum systems such as quantum dots in terms of statistics of detected signals known as full-counting statistics. Quantum gas microscopy enables one to observe many-body systems at the single-atom precision. We extend the idea of full-counting statistics to nonequilibrium open many-particle dynamics and apply it to discuss the quench dynamics. By way of illustration, we consider an exactly solvable model to demonstrate the emergence of unique phenomena such as nonlocal and chiral propagation of correlations, leading to a concomitant oscillatory entanglement growth. We find that correlations can propagate beyond the conventional maximal speed, known as the Lieb-Robinson bound, at the cost of probabilistic nature of quantum measurement. These features become most prominent at the real-to-complex spectrum transition point of an underlying parity-time-symmetric effective non-Hermitian Hamiltonian.  A possible experimental realization with quantum gas microscopy is discussed.
\end{abstract}
\maketitle

The last two decades have witnessed remarkable developments in the ability to detect individual quanta. In small nanoscale devices such as quantum dots, the exchange of electrons with the reservoir has been detected at the single-electron level \cite{WL03,TF04,JB05,GS06}. Photons  emitted from atoms or molecules are now routinely detected individually over a broad range of frequencies \cite{MDE11}. In these systems, complete information about the underlying nonequilibrium dynamics can be obtained from the full-counting statistics \cite{LL96,BW01,KS08,EM09}, i.e., statistics of the number of detected signals. While related techniques were applied to Bose gases \cite{EA04,PA06,SH07} and electron leads \cite{KL09}, developments in this direction have been made for quantum objects with relatively small degrees of freedom.

Meanwhile, recent advances in quantum gas microscopy  \cite{BWS09,SJF10,MM15,CLW15,PMF15,EH15,OA15,EGJA15,RY16,AA16} have enabled one to detect atoms trapped in an optical lattice at the single-atom precision.  Already a number of groundbreaking experiments such as direct observations \cite{CM12,LT13,FT15} of  light-cone spreading of correlations limited by the Lieb-Robinson (LR) velocity $v_{\rm LR}$ \cite{LE72,BS06,NB06}, and measurements of entanglement entropy \cite{IR15} and antiferromagnetic correlations \cite{PM16f,BM16,CLW16f,MA16} have been achieved. 
Similar techniques are available in trapped ions \cite{JP14,RP14}.
On another front, various types of controlled dissipation have been  realized in quantum gases \cite{PD12,PD13,BG13,RB16,LHP17,LJ16,PYS15,TT17}. These  developments suggest possibilities of measuring {\it open many-body} systems at the single-quantum level. 

The aim of this Letter is to extend the idea of full-counting statistics to nonequilibrium many-particle dynamics. We consider a many-particle system coupled to a Markovian reservoir, and discuss the full-counting dynamics that gives the time evolution of the density matrix conditioned on the number of quantum jumps. A quantum jump refers to a discrete stochastic event due to the action of a jump operator known as the Lindblad operator $\hat{L}_{a}$ \cite{SM1}. Physically, this associates with detection of a specific measurable signal. Depending on each realization of quantum jumps, the system evolves in time stochastically (referred to as ``trajectory"). Trajectories can then be classified according to the number of jumps.
We find nonlocal and chiral propagation of correlations, and a concomitant oscillatory entanglement growth. These  features originate from the non-Hermiticity of the underlying open quantum dynamics and  become most prominent at the spectrum transition point of the parity-time ($\mathrm{PT}$) symmetric Hamiltonian \cite{CMB98}. We also discuss a possible experimental realization by quantum gas microscopy.
  
From a broader perspective, previous studies on open quantum  dynamics have revealed emergent thermodynamic structures \cite{GJP10,FC13}, entanglement preparation \cite{LTE14e,MKP14,WAC15,ETJ15}, unconventional phase transitions \cite{LTE14,DS16,YA16}, stochastic dynamics \cite{LMD14,DS15A,YA15,ETJ16A,KW17,YA17,SL17}, topological phenomena \cite{KK18,ZG18}, and reservoir engineering in dissipative systems \cite{BA00,FV09,WY12,HP13,KS14,ETJ16,JC16}. Our work addresses an as yet unexplored question on propagation of correlations and that of information under measurement backaction. Our results indicate that, by harnessing backaction due to observation of individual quanta, correlations can propagate beyond the LR bound at the cost of the probabilistic nature of quantum measurement.

\paragraph{Full-counting many-particle dynamics.\,---} 
We first illustrate our idea in a general way and then apply it to an exactly solvable model. Suppose that a quantum many-particle system is coupled to a Markovian reservoir and described by 
\eqn{\label{masterg}
\frac{d\hat{\rho}(t)}{dt}=-i\left(\hat{H}_{\rm eff}\hat{\rho}-\hat{\rho}\hat{H}_{\rm eff}^{\dagger}\right)+{\cal J}[\hat{\rho}],
}
where $\hat{\rho}(t)$ is the density matrix, $\hat{H}_{\rm eff}=\hat{H}-(i/2)\sum_{a}\hat{L}^{\dagger}_{a}\hat{L}_{a}$ is an effective non-Hermitian Hamiltonian with $\hat{L}_{a}$ being Lindblad operators, and ${\cal J}[\hat{\rho}]=\sum_{a}\hat{L}_{a}\hat{\rho}\hat{L}^{\dagger}_{a}$ describes quantum jump processes \cite{DR92,DJ92,HC93,AD15,SM1}. Here and henceforth we set $\hbar=1$. Given an observed number $n$ of quantum jumps, we consider the full-counting many-particle dynamics described by the density matrix
\eqn{\label{conditional}
\hat{\rho}^{(n)}(t)=\frac{\hat{\cal P}_{n}\hat{\rho}(t)\hat{\cal P}_{n}}{P_{n}(t)},
}
where $\hat{\cal P}_{n}$ is a projector onto the subspace corresponding to $n$ jumps and $P_{n}(t)={\rm Tr}[\hat{\cal P}_{n}\hat{\rho}(t)\hat{\cal P}_{n}]$ gives the probability of finding $n$ jumps during the time interval $[0,t]$. In this Letter, the jump process is assumed to be destructive, i.e., $\hat{L}_{a}$ causes loss of a single particle. In practice, one can obtain $\hat{\rho}^{(n)}(t)$ by initially preparing $N$ particles, letting the system evolve during time $t$, and performing a global measurement to count the total number of particles. Note that a time record of quantum jumps must not be known here. In experiments, similar postselective operations  have found important applications  in ultracold atoms \cite{EM11,FT15,BI08,IR15} and linear optics \cite{KP07}. 

Decomposing the density matrix into the sum $\hat{\rho}=\sum_{n=0}^{N}\hat{\varrho}^{(n)}$ of unnormalized conditional density matrices $\hat{\varrho}^{(n)}=\hat{\cal P}_{n}\hat{\rho}\hat{\cal P}_{n}$, the time evolution can formally be solved as 
\eqn{\label{fc}
\hat{\varrho}^{(n)}(t)&=&\sum_{\{a_{k}\}_{k=1}^{n}}\int_{0}^{t}dt_{n}\cdots\int_{0}^{t_{2}}dt_{1}\prod_{k=1}^{n}\left[\hat{\cal U}_{\rm eff}(\Delta t_{k})\hat{L}_{a_{k}}\right]\nonumber\\
&\times&\hat{\cal U}_{\rm eff}(t_{1})\hat{\rho}(0)\hat{\cal U}_{\rm eff}^{\dagger}(t_{1})\prod_{k=1}^{n}\left[\hat{L}_{a_{k}}^{\dagger}\hat{\cal U}^{\dagger}_{\rm eff} (\Delta t_{k})\right],
} 
where $\Delta t_{k}=t_{k+1}-t_{k}$ with $t_{n+1}\equiv t$, and $\hat{\cal U}_{\rm eff}(t)=e^{-i\hat{H}_{\rm eff}t}$. The solution~(\ref{fc})  represents the ensemble average over all possible occurrences of  $n$ quantum jump events.

 As in closed systems  \cite{PC05,PC06,KC07,CM15,KH13,KM17,FM08,EV08,RG14}, for unconditional open dynamics $\hat{\rho}(t)=\sum_{n}P_{n}(t)\hat{\rho}^{(n)}(t)$, the speed at which correlations build up between distant particles is known to be bounded by the LR velocity \cite{PD10,KM14,SM1}, provided that the Liouvillian  of Eq.~(\ref{masterg}) consists of local operators. In contrast, for the full-counting dynamics $\hat{\rho}^{(n)}(t)$, the propagation speed is no longer expected to obey the LR velocity due to the nonlocal nature of the measurement that acts on an entire many-particle system. Here we explore such hitherto unexplored nonequilibrium dynamics.

\begin{figure}[b]
\includegraphics[width=86mm]{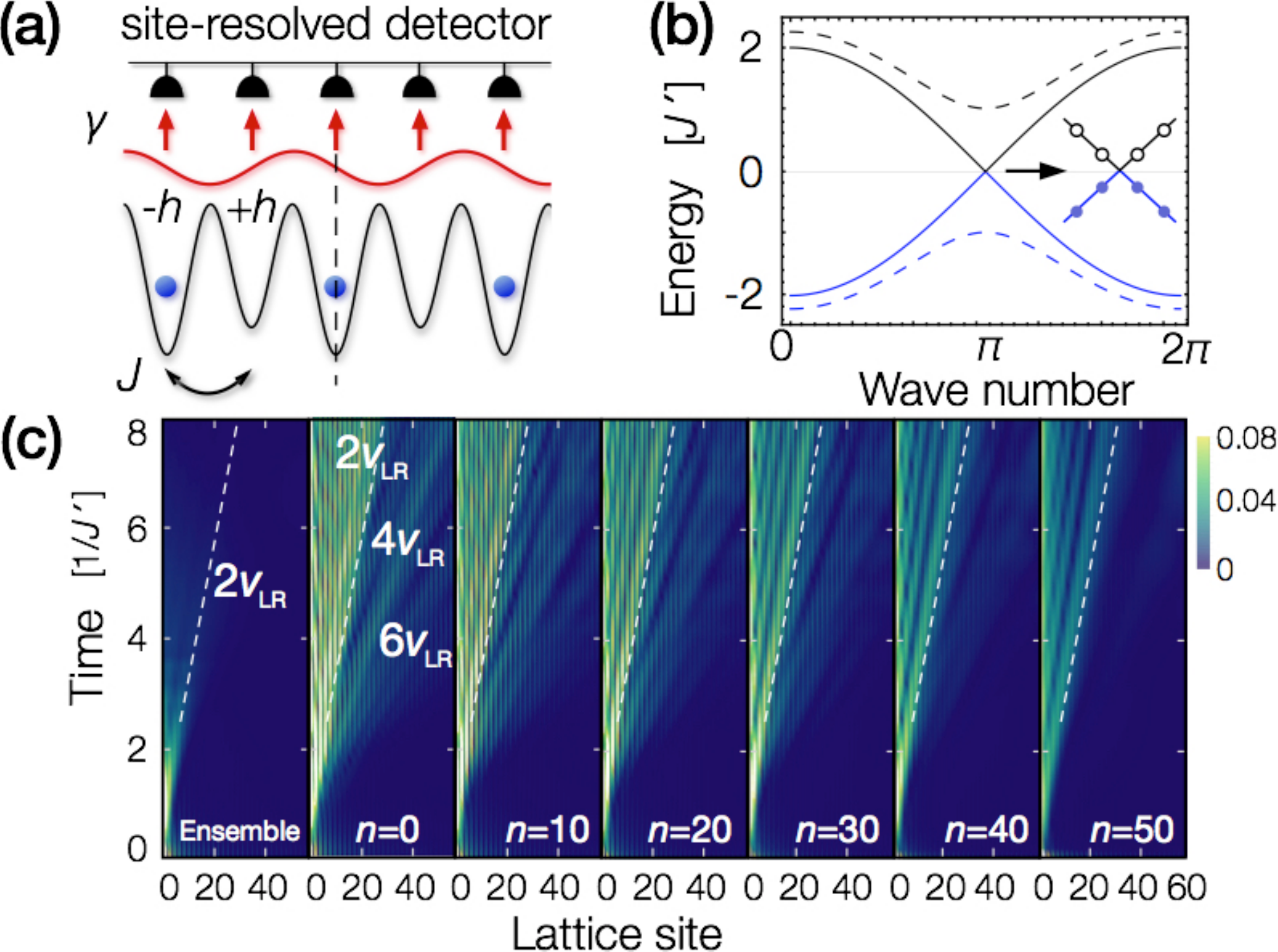} 
\caption{\label{fig1}
(a) Fermions trapped in a superlattice and subject to a spatially modulated dissipative lattice (red), which causes atomic loss and breaks the parity symmetry with respect to the dashed line. The total atom number is measured by site-resolved measurement. (b) Top (black) and bottom (blue) bands of the effective Hamiltonian (\ref{PT}) for $\gamma=0,h=J$ (dashed curves) and $\gamma/h=1/2$ (solid curves). (c) Unconditional (left-most panel) and full-counting (other panels) equal-time correlations for different numbers $n$ of quantum jumps with $N=L/2=61$ and $\gamma=h/2=0.5J$. White dashed lines represent the light cone associated with the Lieb-Robinson bound.
}
\end{figure}

To be concrete, we focus on a simple exactly solvable model. Consider spin-polarized $N$ fermionic atoms trapped in a superlattice with the Hamiltonian $\hat{H}=-\sum_{l=0}^{L-1}[J(\hat{c}_{l+1}^{\dagger}\hat{c}_{l}+\hat{c}_{l}^{\dagger}\hat{c}_{l+1})+(-1)^{l}h\hat{c}_{l}^{\dagger}\hat{c}_{l}]$. Here $\hat{c}_{l}$ ($\hat{c}_{l}^{\dagger}$) is the annihilation (creation) operator of a spinless fermion at site $l$, $J$ is the hopping amplitude, and $h$ describes the on-site staggered potential. We assume that $L$ is even and that the system is initially half-filled, i.e., $N=L/2$. The system is subject to periodic boundary conditions and spatially periodic dissipation which can be induced by a weak resonant optical lattice (Fig.~\ref{fig1}(a)). The time evolution is then described  by the master equation~(\ref{masterg}) 
with the jump process  $\mathcal{J}[\hat{\rho}]=2\gamma\sum_{l}[2\hat{c}_{l}\hat{\rho}\hat{c}_{l}^{\dagger}+(-1)^{l}(\hat{c}_{l}\hat{\rho}\hat{c}_{l+1}^{\dagger}+\hat{c}_{l+1}\hat{\rho}\hat{c}_{l}^{\dagger})]$ and the effective  Hamiltonian $\hat{H}_{\rm eff}=\hat{H}_{\mathrm{PT}}-2i\gamma\hat{N}$. Both of them consist of local operators and the resulting non-Hermitian Hamiltonian

\eqn{\label{PT}
\hat{H}_{\mathrm{PT}}\!&=&\!-\!\!\sum_{l=0}^{L-1}\!\left[\left(J\!+\!(-1)^{l}i\gamma\right)\!\!\left(\hat{c}_{l+1}^{\dagger}\hat{c}_{l}\!+\!\hat{c}_{l}^{\dagger}\hat{c}_{l+1}\right)\!+\!(-1)^{l}h\hat{c}_{l}^{\dagger}\hat{c}_{l}\right]
\nonumber\\
&=&\sum_{0\leq k<2\pi}\sum_{\lambda=\pm}\epsilon_{\lambda}(k)\hat{g}_{\lambda k}^{\dagger}\hat{f}_{\lambda k}
}
satisfies the $\mathrm{PT}$ symmetry \cite{CMB98}, i.e., the symmetry with respect to the product of parity operation and time reversal. The jump part can be written in the diagonal form $\hat{L}_{a}=\sqrt{\gamma_{\lambda k}}\hat{d}_{\lambda k}$ with $\gamma_{\lambda k}$ being positive coefficients and $\hat{d}_{\lambda k}$ being a linear combination of $\hat{c}_{l}$ \cite{SM1}. Physically, this jump operator annihilates a single-particle mode with wavevector $k$.

In the last line of Eq.~(\ref{PT}), the effective Hamiltonian $\hat{H}_{\mathrm{PT}}$ is diagonalized with eigenvalues $\epsilon_{\pm}(k)=\pm\sqrt{h^2-4\gamma^2+2J'^2(1+\cos(k))}$, where $J'=\sqrt{J^2+\gamma^2}$ and $k=2\pi n/(L/2)$ ($n=0,1,\ldots,L/2-1$). The operators $\hat{g}^{\dagger}$ and $\hat{f}$ create the right and left eigenvectors, i.e., $\hat{H}_{\mathrm{PT}}\hat{g}^{\dagger}_{\lambda k}|0\rangle=\epsilon_{\lambda}(k)\hat{g}^{\dagger}_{\lambda k}|0\rangle$ and $\langle 0|\hat{f}_{\lambda k}\hat{H}_{\mathrm{PT}}=\langle 0|\hat{f}_{\lambda k} \epsilon_{\lambda}(k)$, and they obey a generalized anticommutation relation $\{\hat{f}_{\lambda k},\hat{g}^{\dagger}_{\lambda' k'}\}=\delta_{k,k'}\delta_{\lambda,\lambda'}$. A direct consequence of the non-Hermiticity is the nonorthogonality of eigenvectors. Specifically, $\hat{g}$ and $\hat{g}^{\dagger}$ satisfy an unusual commutation relation $\{\hat{g}_{\lambda k},\hat{g}^{\dagger}_{\lambda' k'}\}=\delta_{k,k'}\Delta_{\lambda\lambda'}(k)$,
where $\Delta_{\lambda\lambda'}(k)$ is the $2\times2$ matrices whose nonzero off-diagonal elements indicate the nonorthogonality between the right eigenvectors of different bands in mode $k$. 

When $\gamma<h/2$, $\hat{H}_{\mathrm{PT}}$ has an entirely real, gapped band spectrum (dashed curves in Fig.~\ref{fig1}(b)). At $\gamma=h/2$, the band gap closes at $k=\pi$ (solid curves in Fig.~\ref{fig1}(b)), and the $k=\pi$ eigenstates of the two bands coalesce into a single one. Such a point is known as an exceptional point \cite{TK80} or, in the thermodynamic limit, as the spectral singularity \cite{AM09}. Above the threshold $\gamma>h/2$, some eigenmodes around $k=\pi$ turn out to have complex pairs of pure imaginary eigenvalues.  
For simplicity, we assume that $L/2$ is odd such that the singularity at $k=\pi$ is avoided (see the inset in Fig.~\ref{fig1}(b)) \footnote{For an even $L/2$, one can impose the anti-periodic boundary conditions to ensure that $k=\pi$ is avoided.}. 
 
\paragraph{Nonlocal propagation of correlations.\,---} 
Combining a general solution~(\ref{fc}) and the diagonalized effective Hamiltonian~(\ref{PT}), we obtain an exact solution of the full-counting dynamics $\hat{\rho}^{(n)}(t)$ \cite{SM1}. By way of illustration, we consider the following quench dynamics. Initially, the staggered potential $h$ and the dissipation $\gamma$ are switched off and the system is prepared in the ground state of $\hat{H}$. We then suddenly switch on $h$ and $\gamma$, and let the system evolve according to Eq.~(\ref{masterg}). We choose $\gamma=h/2$ such that the parameters of the postquench Hamiltonian $\hat{H}_{\mathrm{PT}}$ are set to the real-to-complex spectrum transition point, leading to the linear dispersion around $k=\pi$ (Fig.~\ref{fig1}(b)).

Let us first discuss the unconditional case $\hat{\rho}(t)=\sum_{n=0}^{N}P_{n}(t)\hat{\rho}^{(n)}(t)$.  The left-most panel in Fig.~\ref{fig1}(c) plots an equal-time correlation $C(l,t)={\rm Tr}[\hat{\rho}(t)\hat{c}^\dagger_{l}\hat{c}_{0}]$, which exhibits a blurred light cone \cite{BJS17}. Since the Liouvillian of Eq.~(\ref{masterg}) consists of local operators, it is expected that correlations can propagate no faster than twice the LR velocity $2v_{\rm LR}$ \cite{BS06,PD10,KM14,SM1}, where $v_{\rm LR}$ is given by the maximum group velocity $|\partial\epsilon_{\pm}(k)/\partial (k/2)|_{k=\pi}=2J'$ \footnote{Here a factor of two in the group velocity comes from our choice of wavevectors defined in the Fourier transforms of the sublattices (see Eq.~(\ref{neqcorr})).}.

The situation is quite different in the full-counting dynamics $\hat{\rho}^{(n)}(t)$ in Eq.~(\ref{conditional}). Figure~\ref{fig1}(c) plots an equal-time correlation 
$C^{(n)}(l,t)={\rm Tr}[\hat{\rho}^{(n)}(t)\hat{c}^\dagger_{l}\hat{c}_{0}]$ for such dynamics with different values of $n$. We find nonlocal modes that propagate faster than the LR velocity of the corresponding unconditional dynamics. Moreover, velocities of such supersonic modes appear at integer multiples of $2v_{\rm LR}$. Physically, the propagations beyond the LR bound signals nonlocality encoded in the full-counting dynamics $\hat{\rho}^{(n)}(t)$. 

The origin of these nonlocal propagations can be understood from the underlying dynamics governed by the effective non-Hermitian Hamiltonian $\hat{H}_{\mathrm{PT}}$, which describes time evolution during an interval without quantum jumps \cite{SM1}. 
To clarify the essential point, let us focus on a simple quantum trajectory containing null jumps: $\hat{\rho}^{(0)}(t)=e^{-i\hat{H}_{\mathrm{PT}}t}\hat{\rho}(0)e^{i\hat{H}_{\mathrm{PT}}^{\dagger}t}/{\rm Tr}[e^{-i\hat{H}_{\mathrm{PT}}t}\hat{\rho}(0)e^{i\hat{H}_{\mathrm{PT}}^{\dagger}t}],$
where the factor $-2i\gamma\hat{N}$ in $\hat{H}_{\rm eff}$ cancels out in forming the ratio. A similar time evolution has been discussed in dissipative evolutions \cite{NG81} and $\mathrm{PT}$-symmetric quantum systems \cite{BDC12,KK17}. 
An initially pure state remains pure in this dynamics \cite{BDC12}. Denoting $\hat{\rho}(0)=|\Psi_{0}\rangle\langle\Psi_{0}|$, we introduce an unnormalized time-dependent wavefunction $|\Psi_{t}\rangle=e^{-i\hat{H}_{\mathrm{PT}}t}|\Psi_{0}\rangle$. We first expand the initial state $|\Psi_{0}\rangle$ in terms of right eigenvectors: $|\Psi_{0}\rangle=\prod_{k}[\sum_{\lambda}\psi_{\lambda k}\hat{g}_{\lambda k}^{\dagger}]|0\rangle$, where $\psi_{\lambda k}$'s are expansion coefficients. 
We then introduce the unequal-time correlation by $\tilde{C}^{(0)}(l,t)=\langle\Psi_{0}|\hat{c}_{l}^{\dagger}(t)\hat{c}_{0}(0)|\Psi_{0}\rangle/\langle\Psi_{t}|\Psi_{t}\rangle$ with $\hat{c}^{\dagger}_{l}(t)=e^{i\hat{H}_{\rm PT}^{\dagger}t}\hat{c}^{\dagger}_{l}e^{-i\hat{H}_{\rm PT}t}$, which can be calculated as 
\eqn{\label{neqcorr}
\!\!\!\!\tilde{C}^{(0)}(l,t)\!=\!\frac{2}{L}\!\sum_{k}\sum_{\lambda=\pm}\begin{Bmatrix}\alpha_{\lambda k}\\ \beta_{\lambda k}\end{Bmatrix}\frac{\psi_{\lambda k}^{*}e^{i\epsilon_{\lambda}(k)t-ik\lceil l/2\rceil}}{{\cal N}_{k}(t)}.
}
Here $\alpha_{\lambda k}$ and $\beta_{\lambda k}$  are coefficients chosen according to the parity of $l$ \cite{SM1}, $\lceil \cdot\rceil$ is the ceiling function, and ${\cal N}_{k}(t)=\sum_{\lambda\lambda'=\pm}\psi^{*}_{\lambda k}(t)\Delta_{\lambda\lambda'}(k)\psi_{\lambda' k}(t)$, where $\psi_{\lambda k}(t)=\psi_{\lambda k}e^{-i\epsilon_{\lambda}(k)t}$. The total norm of an unnormalized quantum state $|\Psi_{t}\rangle$ is then given by $\langle\Psi_{t}|\Psi_{t}\rangle=\prod_{k}{\cal N}_{k}(t)$. 

\begin{figure}[t]
\includegraphics[width=86mm]{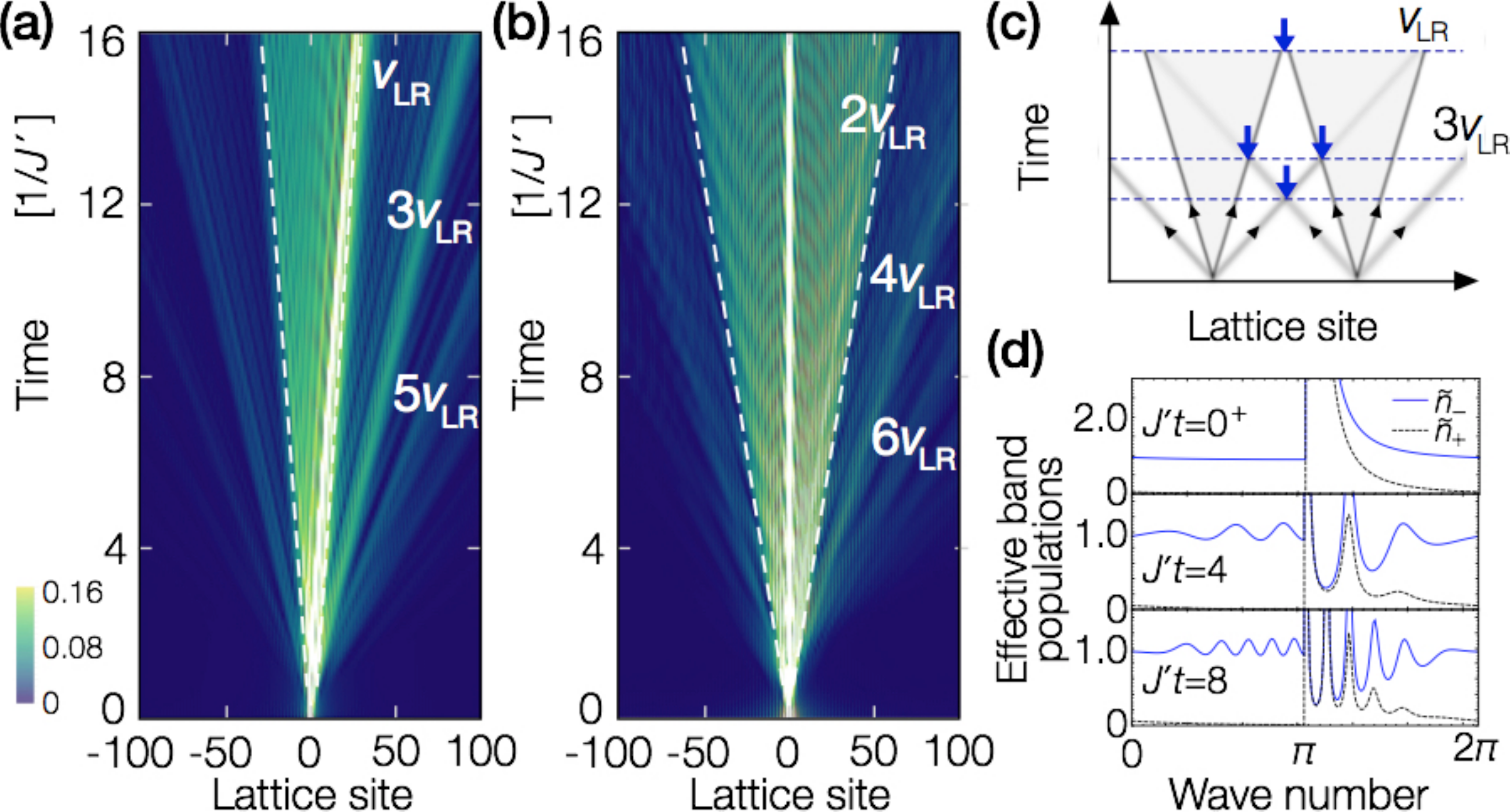} 
\caption{\label{fig2}
(a) Unequal- and (b) equal-time correlations plotted for $N=L/2=61$ and $\gamma=h/2=0.5J$. The white dashed lines indicate the Lieb-Robinson bound. (c) Illustration of how the correlation is carried by quasiparticles propagating at velocities $v_{\rm LR}$ and $3v_{\rm LR}$. (d) Effective band populations for different times $J't=0^+$ (postquench state), $4,$ and $8$. 
}
\end{figure}

A crucial observation here is that due to the nonorthogonality of  eigenvectors ($\Delta_{+-}=\Delta_{-+}^{*}\neq 0$) the norm ${\cal N}_{k}(t)$ oscillates at frequency $2\epsilon_{+}(k)$. Thus, $\tilde{C}^{(0)}(l,t)$ in Eq.~(\ref{neqcorr}) involves terms that oscillate at frequencies $\epsilon_{\lambda}(k),3\epsilon_{\lambda}(k),5\epsilon_{\lambda}(k),\ldots$, leading to the propagations at velocities $v_{\rm LR},3v_{\rm LR},5v_{\rm LR},\ldots$ (Fig.~\ref{fig2}(a)). In contrast, the equal-time correlation $C^{(0)}(l,t)$ involves the propagations at velocities  $2v_{\rm LR},4v_{\rm LR},6v_{\rm LR},\ldots$ (Fig.~\ref{fig2}(b)), as it is formed by quasiparticle pairs propagating with velocities $v_{\rm LR},3v_{\rm LR},5v_{\rm LR},\ldots$ \cite{PC05} (Fig.~\ref{fig2}(c)). 

The emergence of these supersonic modes is a consequence of the interplay between non-Hermiticity and the many-particle nature of the system. The appearance of the oscillating norm factors ${\cal N}_{k}(t)$ in the denominator in Eq.~(\ref{neqcorr}) originates from the fact that the total norm of a  many-particle quantum state is given by their product.
Thus, the supersonic modes have no counterparts in the single-particle sector or the mean-field non-Hermitian dynamics in optics \cite{CER10,AR12} and dissipative matter waves \cite{BG13,PP16,GT15,KVV16}, where the total norm is determined by the sum rather than the product of ${\cal N}_{k}(t)$ \cite{KGM08}. 

In analogy with closed systems \cite{PC05,PC06,KC07,CM15,KH13,KM17,FM08,EV08,RG14}, we may regard $\tilde{n}_{\lambda k}(t)=|\psi_{\lambda k}|^2/{\cal N}_{k}(t)$ as an effective band population of quasiparticles. In the noninteracting closed system, the band population remains constant after the quench \cite{PC05,FM08,EV08,RG14}. In contrast, the effective band population oscillates in time (Fig.~\ref{fig2}(d)) due to the nonorthogonality between two eigenvectors in a mode $k$. Since the interference for the same momentum implies a nonlocal coupling in real space, we may interpret the supersonic propagation as a consequence of such a nonlocal, self-interaction of quasiparticles.

\begin{figure}[t]
\includegraphics[width=86mm]{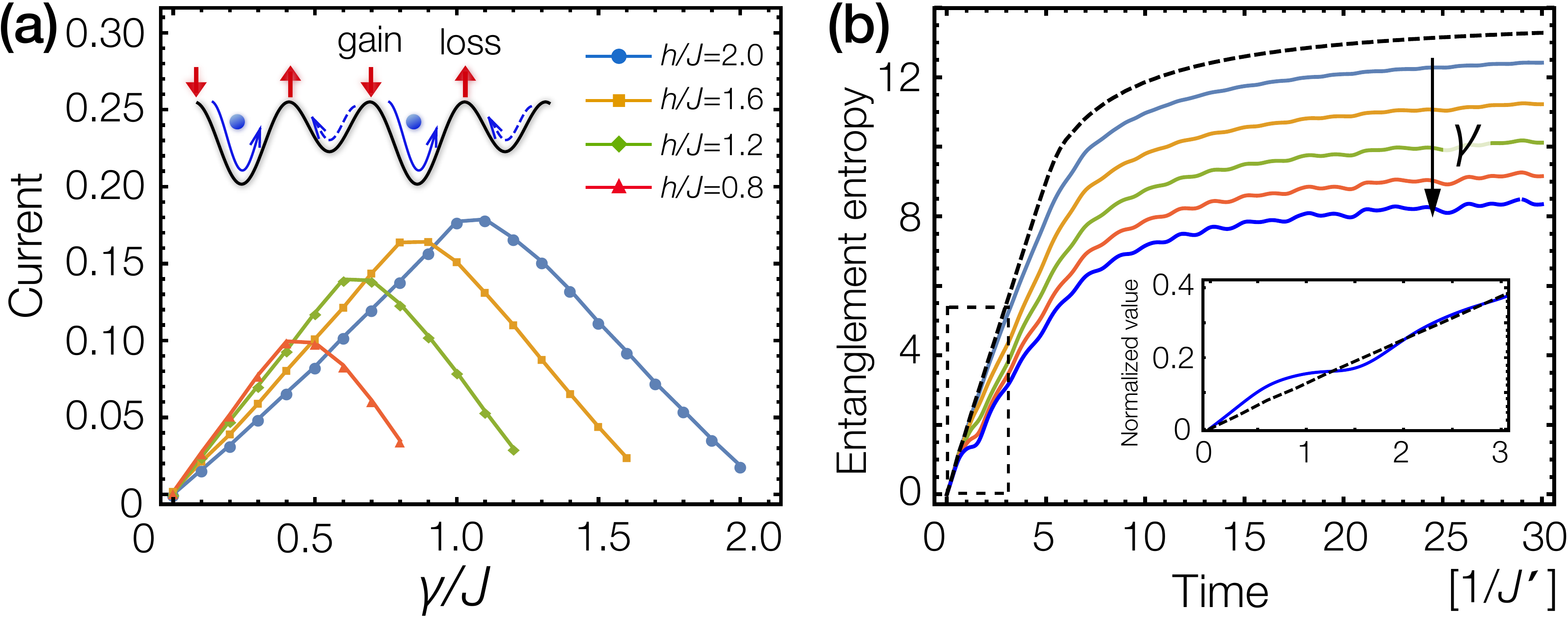} 
\caption{\label{fig3}
(a) Current averaged over the time interval between $t=15/J'$ and $20/J'$ plotted against the strength of dissipation $\gamma$ for different on-site potentials $h$. The inset shows a gain-loss situation for a positive particle current. (b) Time evolutions of entanglement entropy of a chain of length $20$ starting from a product state (the ground state of $\hat{H}$ with $h=\infty$). We vary a postquench parameter $\gamma$ from $0.0$ to $0.5$ (top to bottom) with step $0.1$ and $\gamma=h/2$ held fixed. The inset magnifies the short-time regime showing the oscillatory behavior due to the time-dependent effective band populations.
}
\end{figure}

\paragraph{Chirality in propagation of correlations.\,---} 
Yet another feature of the observed propagation is its chirality. Here by chirality we mean that the violation of the left-right symmetry of propagation of correlations. This symmetry breaking results from the parity violation in the effective Hamiltonian, i.e., $\hat{H}_{\rm PT}$ is not invariant under $l\to -l$ (Fig.~\ref{fig1}(a)).
We can intuitively interpret the pronounced propagation in the right direction found in Figs.~\ref{fig2}(a,b) on the basis of the gain-loss structure of $\hat{H}_{\rm PT}$. Imagine that particles are injected at the ``gain" bond having positive imaginary hopping $+i\gamma$ (inset of Fig.~\ref{fig3}(a)). Then, a majority of the particles flow into the deeper, right potential. The injected particles are removed at the ``loss" bond and thus local  flows of particles can be  formed. Overall, the flow in the right direction overweighs the reverse flow, resulting in a net positive current. 

A nontrivial feature here is that the chirality is most pronounced at the spectrum transition point of $\hat{H}_{\mathrm{PT}}$. Figure~\ref{fig3}(a) shows the current $iJ\sum_{l=0}^{L-1}(\hat{c}_{l}^{\dagger}\hat{c}_{l+1}-\hat{c}_{l+1}^{\dagger}\hat{c}_{l})$ in a long-time regime for different values of $h$ and $\gamma$. The pronounced chirality at the threshold $\gamma=h/2$ originates from the emergence of the exceptional point at $k=\pi$ (Fig.~\ref{fig1}(b)) \cite{HWD01}. In its vicinity, the strong nonorthogonality induces coalescence of two eigenvectors of different bands into the one associated with the band dispersion having positive group velocities $\partial \epsilon/\partial (k/2)>0$ \cite{SM1}. 
This confluent band structure leads to imbalanced effective band populations $\tilde{n}_{\lambda k}(t)$ in Fig.~\ref{fig2}(d), where the population $\tilde{n}_{+,k}$ of the upper band almost vanishes for $k<\pi$, while $\tilde{n}_{-,k}$ takes a value close to unity. Such an effective violation of the particle-hole symmetry generates quasiparticles having positive group velocities (c.f.~Fig.~\ref{fig1}(b)), leading to the pronounced propagation of correlations in the right direction. 

It is noteworthy that, in contrast to the single-particle sector \cite{LS09,LSS10,LZ11,SD12,LF13,WJ14}, the chirality in the present case is prominent owing to the formation of the Fermi sea at $\epsilon_{\rm F}=0$; low-energy excitations are subject to the strong nonorthogonality that becomes maximal at the gap closing point $k=\pi$.  The resulting unidirectionality appears as the chiral propagation of correlations in the case of many-particle systems.

The chirality also has a physical consequence in the entanglement growth of the system. Figure~\ref{fig3}(b) shows the time evolution of the entanglement entropy $S_{A}[\hat{\rho}^{(0)}(t)]$ \cite{IP03} after the quench for different values of $\gamma$ with subregion $A$ of a chain of 20 sites. A decrease in the entanglement entropy with increasing $\gamma$ can be interpreted as a consequence of the chirality. 
Quantum quench generates pairs of entangled quasiparticles propagating in opposite directions \cite{PC05}. The entanglement entropy essentially measures the number of quantum-mechanically correlated pairs such that one quasiparticle is inside and the other is outside of subregion $A$. Since the chiral (unidirectional) modes do not generate such entangled pairs moving in opposite directions, the chirality leads to a decrease in the entanglement entropy. 

An oscillation on top of a linear increase in the entanglement entropy (inset in Fig.~\ref{fig3}(b)) results from the time-dependent effective band populations $\tilde{n}_{\lambda k}(t)$. In view of the simple dispersion of the present model, this presents yet another unique feature of open quantum dynamics because, in closed integrable systems, such oscillations of entanglement entropy emerge only if there exist multiple local maxima in a band dispersion \cite{FM08}. 

\paragraph{Discussions.\,---} As a possible experimental test of the present consideration, we propose to using site-resolved measurements \cite{BWS09,SJF10,MM15,CLW15,PMF15,EH15,OA15,EGJA15,RY16,AA16} to probe the full-counting dynamics. The dissipation can be implemented by superimposing a weak resonant optical lattice \cite{RB16,LHP17,LJ16,PP16,MKO99,AT03} (see Fig.~\ref{fig1}(a)). The parameters $\gamma,J,h$ are experimentally tuned by controlling the intensities of optical beams. Using fermionic quantum gas microscopy \cite{CLW15,PMF15,EH15,OA15,EGJA15}, one can measure the site-resolved density-density correlation and the total number of particles simultaneously. Since the connected density-density correlation in noninteracting models reduces to the product of the equal-time correlations, both correlations share the same information. 
While detecting supersonic propagations will be challenging at long times, they should be observable in a short-time regime such that a relatively large number of atoms still remain in the trap. For example, if one chooses $\gamma=h/2=0.25J$ and $N=61$, the probability of detecting trajectories with lost particles less than the half of the initial total particle number can exceed $\sim\!\!20\%$ up to $tJ'\lesssim 3$ at which supersonic propagations similar to the ones in Fig.~\ref{fig1}(c) are visible. In practice, one may choose $^{6}$Li atoms and use an optical beam resonant with the $^{2}S_{1/2}\!\to$ $\!^{2}P_{3/2}$ transition as recently demonstrated in Ref.~\cite{LJ16}.

The ability of measuring individual quanta can reveal the emergence of unique many-particle dynamics that cannot be seen in closed systems. Our results show that correlations can propagate faster than the LR bound at the cost of the probabilistic nature of quantum measurement. The emergence of the nonlocal propagation originates from the nonorthogonality of eigenvectors due to the non-Hermiticity of the underlying dynamics. In view of the generality of nonorthogonality in non-Hermitian systems, the nonlocal propagation can also appear in a variety of other open many-particle systems. Such features will become most prominent when nonorthogonality becomes maximal due to, for example, the presence of an exceptional point as demonstrated in our paper.
It is intriguing to explore roles of interactions \cite{KC07,CM15} or nonintegrablility \cite{KH13,KM17} in such unconventional many-body dynamics subject to single-quantum resolved measurement. Analogous to closed systems \cite{PC05,PC06}, it is of interest to develop field-theoretic arguments. It is noteworthy that low-energy field theory \cite{YA17nc} of the effective Hamiltonian $\hat{H}_{\mathrm{PT}}$ corresponds to quantum Liouville theory which attracts much attention in high-energy physics \cite{SN91}. We hope that the present work stimulates further studies in these directions.

We are grateful to Juan P. Garrahan, Peter Rabl, Vladimir Konotop, Yogesh Joglekar, Naomichi Hatano, J{\"o}rg Schmiedmayer, Shunsuke Furukawa, Hosho Katsura, Takashi Mori, Keiji Saito, and Zala Lenarcic for valuable discussions. We acknowledge support from KAKENHI Grant No.~JP26287088 from the Japan Society for the Promotion of Science (JSPS), and a Grant-in-Aid for Scientific Research on Innovative Areas ``Topological Materials Science" (KAKENHI Grant No.~JP15H05855), and the Photon Frontier Network Program from MEXT of Japan, ImPACT Program of Council for Science, Technology and Innovation (Cabinet Office, Government of Japan). Y.~A. acknowledges support from JSPS (Grant No.~JP16J03613). 

\bibliography{reference}

\widetext
\pagebreak
\begin{center}
\textbf{\large Supplementary Materials}
\end{center}

\renewcommand{\theequation}{S\arabic{equation}}
\renewcommand{\thefigure}{S\arabic{figure}}
\renewcommand{\bibnumfmt}[1]{[S#1]}
\setcounter{equation}{0}
\setcounter{figure}{0}

\subsection{Brief summary of the quantum trajectory approach}
The quantum trajectory approach to open quantum systems has been originally developed in the field of quantum optics in parallel by several groups having rather different motivations such as quantum measurement \cite{DR92,HC93} or laser cooling of atoms \cite{DJ92}. It provides an intuitive physical picture of the dynamics of systems subject to continuous observation. The quantum trajectory approach is also important as an efficient numerical method for open quantum systems since it allows one to solve the master equation by taking the ensemble average over stochastic time evolutions of pure quantum states, thus avoiding the complexity of finding the full-density matrix. In this section, we briefly review the quantum trajectory approach from a perspective of quantum measurement.

We consider quantum measurement processes characterized by the following measurement operators:
\eqn{\label{conmeas}
\hat{M}_{0}&=&1-i\left(\hat{H}-\frac{i}{2}\sum_{a=1}^{M}\hat{L}_{a}^{\dagger}\hat{L}_{a}\right)dt\equiv1-i\hat{H}_{{\rm eff}}dt,\\
\hat{M}_{a}&=&\hat{L}_{a}\sqrt{dt}\;\;(a=1,2,\ldots,M),\label{conmeass}
}
where $\hat{M}_{0}$ acts on a quantum state if no  signals labeled by $a=1,2,\ldots,M$ are observed during the time interval $[t,t+dt]$ and $\hat{M}_{a}$ acts on it if a signal labeled by $a$ is detected. Here, $\hat{H}$ is the Hamiltonian governing the unitary dynamics of the system, $\hat{L}_{a}$ is an operator associated with a measurable signal $a$, and $\hat{H}_{\rm eff}=H-(i/2)\sum_{a}\hat{L}_{a}^{\dagger}\hat{L}_{a}$ is an effective non-Hermitian Hamiltonian.  We assume that the initial state is pure and thus the state remains so in the course of time evolution. Note that the measurement operators satisfy the normalization condition (aside from a negligible contribution of the order of $O(dt^2)$):
\eqn{
\sum_{a=0}^{M}\hat{M}_{a}^{\dagger}\hat{M}_{a}=1.
}

The detection of a measurable signal is a stochastic process, reflecting the probabilistic nature of quantum measurement.  Its probability is characterized by the expectation value of the square of the measurement operator $\hat{M}_{a}$ with respect to a quantum state $|\psi\rangle$. In the language of stochastic processes, this is formulated as a discrete random variable $dN_{a}$ having the mean value as follows:
\eqn{
E[dN_{a}]=\langle\psi|\hat{M}_{a}^{\dagger}\hat{M}_{a}|\psi\rangle=\langle\psi|\hat{L}_{a}^{\dagger}\hat{L}_{a}|\psi\rangle dt,
}
where $E[\cdot]$ represents the ensemble average over the stochastic process. 
These random variables are assumed to satisfy the stochastic calculus
\eqn{\label{stN}
dN_{a}dN_{b}=\delta_{ab}dN_{a}.
}
Using these notations, the stochastic change of a quantum state $|\psi\rangle$ in the time interval $[t,t+dt]$ can be obtained as
\eqn{\label{stochastic}
|\psi\rangle\to|\psi\rangle+d|\psi\rangle=\left(1-\sum_{a=1}^{M}E[dN_{a}]\right)\frac{\hat{M}_{0}|\psi\rangle}{\sqrt{\langle\psi|\hat{M}_{0}^{\dagger}\hat{M}_{0}|\psi\rangle}}+\sum_{a=1}^{M}dN_{a}\frac{\hat{M}_{a}|\psi\rangle}{\sqrt{\langle\psi|\hat{M}_{a}^{\dagger}\hat{M}_{a}|\psi\rangle}}.
}
Physically, the first term on the right-hand side describes the no-count process occurring with probability $1-\sum_{a=1}^{M}E[dN_{a}]$ and the second term describes the detection of a measurable signal $a$ occurring with probability $E[dN_{a}]$. The latter process is known as quantum jump process and associates with, for example, the detection of photons or atoms, in which $\hat{L}_{a}$ is an annihilation operator of the photon or atom field. We note that the denominator in each term is introduced to ensure the normalization of the state vector. 

From Eqs.~(\ref{conmeas}) and (\ref{conmeass}), we can rewrite Eq.~(\ref{stochastic}) as
\eqn{\label{conmeas2}
d|\psi\rangle=\left(1-i\hat{H}_{{\rm eff}}+\frac{1}{2}\sum_{a=1}^{M}\langle\psi|\hat{L}_{a}^{\dagger}\hat{L}_{a}|\psi\rangle\right)dt|\psi\rangle+\sum_{a=1}^{M}\left(\frac{\hat{L}_{a}|\psi\rangle}{\sqrt{\langle\psi|\hat{L}_{a}^{\dagger}\hat{L}_{a}|\psi\rangle}}-|\psi\rangle\right)dN_{a}.
}
The first term in the right-hand side describes the non-Hermitian time evolution, in which the factor $\sum_{a}\langle\psi|\hat{L}_{a}^{\dagger}\hat{L}_{a}|\psi\rangle/2$ keeps the normalization of the state vector. In the second term, when a signal $a$ is detected, an operator $\hat{L}_{a}$ acts on the quantum state and causes its discontinuous change (``jump"). In this sense, $\hat{L}_{a}$ is also known as a jump operator. Integrating this stochastic differential equation numerically, one can obtain a realization of the time evolution of a pure quantum state, which is referred to as a quantum trajectory. Taking the ensemble average over all possible trajectories, one can reproduce the Lindblad master equation. To see this explicitly, let us rewrite Eq.~(\ref{conmeas2}) using the density matrix $\hat{\rho}_{p}=|\psi\rangle\langle\psi|$ of a pure state: 

\eqn{\label{conmeas3}
d\hat{\rho}_{p}=-i\left(\hat{H}_{{\rm eff}}\hat{\rho}_{p}-\hat{\rho}_{p}\hat{H}_{{\rm eff}}^{\dagger}\right)dt+\sum_{a=1}^{M}\langle\psi|\hat{L}_{a}^{\dagger}\hat{L}_{a}|\psi\rangle\hat{\rho}_{p}dt+\sum_{a=1}^{M}\left(\frac{\hat{L}_{a}\hat{\rho}_{p}\hat{L}_{a}^{\dagger}}{\langle\psi|\hat{L}_{a}^{\dagger}\hat{L}_{a}|\psi\rangle}-\hat{\rho}_{p}\right)dN_{a},
}
where we take the leading order of $O(dt)$ and use the stochastic calculus~(\ref{stN}).
Introducing the ensemble-averaged density matrix $\hat{\rho}=E[\hat{\rho}_{p}]$ and taking the average of Eq.~(\ref{conmeas3}), one can show that the density matrix $\hat{\rho}$ obeys the Lindblad master equation
\eqn{
\frac{d\hat{\rho}}{dt}=-i\left(\hat{H}_{{\rm eff}}\hat{\rho}-\hat{\rho}\hat{H}_{{\rm eff}}^{\dagger}\right)+\sum_{a=1}^{M}\hat{L}_{a}\hat{\rho}\hat{L}_{a}^{\dagger},
}
which is Eq.~(1) in the main text.

\subsection{Lieb-Robinson bound and its relation to the full-counting dynamics}
Here we briefly summarize the statement and significance of the Lieb-Robinson bound and describe its relation to the full-counting dynamics discussed in the main text. Lieb and Robinson (LR) have shown \cite{LE72} that, for a unitary time evolution in nonrelativistic quantum spins (or fermonic particles for spin-1/2 case) on a lattice, there exists a finite group velocity $v_{\rm LR}$ which bounds the velocity of propagation of information in the system. Specifically, they have shown the bound
\eqn{\label{lrbound}
\left\Vert [\hat{O}_{A}(t),\hat{O}_{B}(0)]\right\Vert \leq c\,{\rm min}(|A|,|B|)\,\left\Vert \hat{O}_{A}\right\Vert \left\Vert \hat{O}_{B}\right\Vert \exp\left({-\frac{L-v_{{\rm LR}}t}{\xi}}\right),
}
where $\hat{O}_{A}$ and $\hat{O}_B$ are local operators acing on two subsystems $A$ and $B$ that are separated by the distance $L$, $\Vert\cdot\Vert$ is the operator norm, $|A|$ ($|B|$) denotes the volume of  $A$ ($B$), $v_{\rm LR}$ is the LR velocity, $\xi$ characterizes the size of the tail in the effective light cone. The operator $\hat{O}(t)$ denotes the Heisenberg representation.  We note that values of constants $c,v_{\rm LR}$ and $\xi$ cannot be given by the bound in general. Physically, the bound~(\ref{lrbound}) shows that a signal given in $B$ at $t=0$ cannot be transferred to $A$ faster than the velocity $v_{\rm LR}$.  Bravyi, Hastings and Verstraete have used the inequality~(\ref{lrbound}) to obtain the bound on the connected equal-time correlation functions after the quench \cite{BS06}:
\eqn{\label{lrbound2}
\langle\Psi_t|\hat{O}_A \hat{O}_B|\Psi_t\rangle<c'(|A|+|B|)\exp\left(-\frac{L-2v_{\rm LR}t}{\chi}\right),
} 
where $c'$ and $\chi$ are constants. These relations play crucial roles especially in quantum information science and have laid the cornerstone in studies of gapped many-body ground states. Later,   the bounds have been generalized \cite{PD10,KM14} to the open quantum dynamics described by the Lindblad master equation, where the Liouvillian is assumed be the sum of local operators acting on the density matrix. This condition is satisfied in our model as inferred from the non-Hermitian term $-\sum_{l}[(-1)^{l}i\gamma(\hat{c}_{l+1}^{\dagger}\hat{c}_{l}+\hat{c}_{l}^{\dagger}\hat{c}_{l+1})+2i\gamma\hat{c}^{\dagger}_{l}\hat{c}_{l}]$ in the effective Hamiltonian and the jump term $\mathcal{J}[\hat{\rho}]=2\gamma\sum_{l}[2\hat{c}_{l}\hat{\rho}\hat{c}_{l}^{\dagger}+(-1)^{l}(\hat{c}_{l}\hat{\rho}\hat{c}_{l+1}^{\dagger}+\hat{c}_{l+1}\hat{\rho}\hat{c}_{l}^{\dagger})]$; both of them consist of only local operators.

\begin{figure}[t]
\includegraphics[width=140mm]{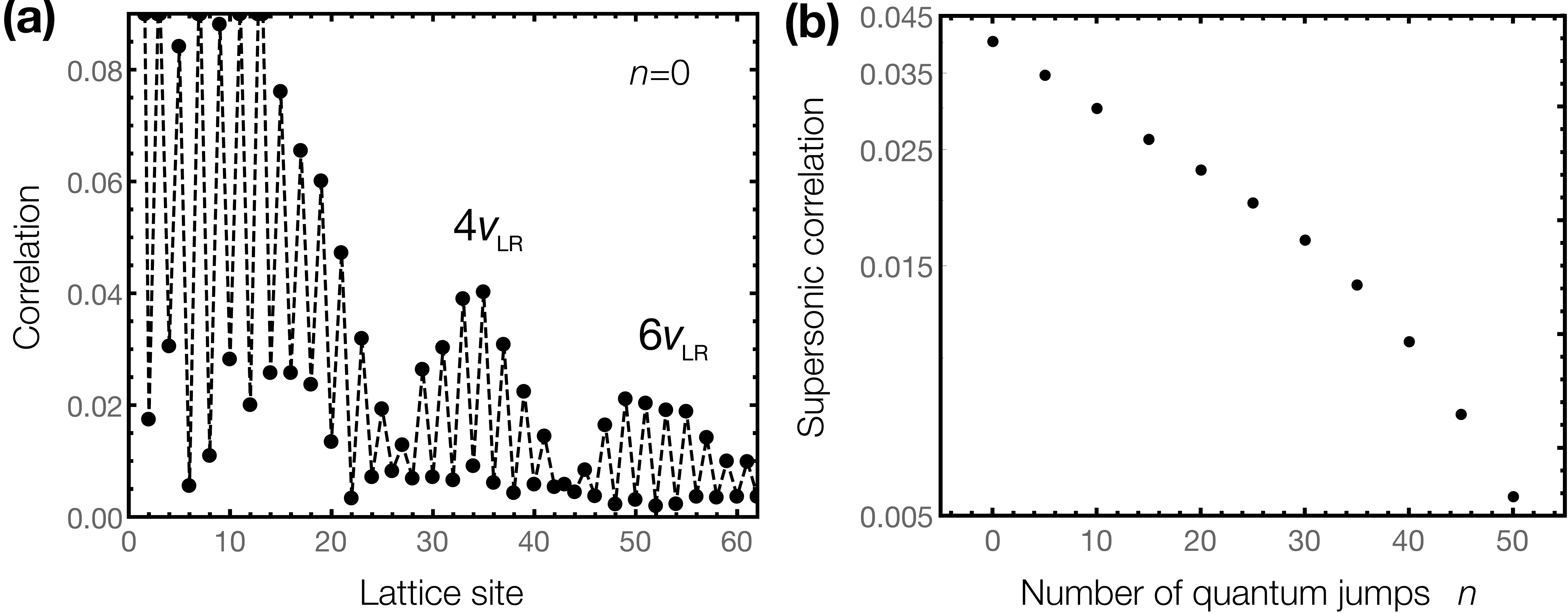}
\caption{\label{figs3}
\noindent{(a) Absolute value of the equal-time correlation $|C^{(n)}(l,t_0)|=|{\rm Tr}[\hat{\rho}^{(n)}(t)\hat{c}_{l}^\dagger\hat{c}_{0}]|$ for the null quantum jump $n=0$ at time $t_0=5/J'$ plotted against  lattice site $l$. All the parameters and the quench protocol are the same as those in Fig. 1(c) in the main text. (b) Equal-time correlation $C^{(n)}(l_0,t_0)$ associated with the supersonic modes propagating with the velocity $4v_{\rm LR}$ plotted against the number of quantum jumps. The lattice site is chosen to be $l_0=35$ at which the supersonic contribution is maximal (see also the panel (a)).}
} 
\end{figure}

In the full-counting dynamics $\hat{\rho}^{(n)}(t)$ (see, e.g., Figs.~2(a) and (b) in the main text), there exist the robust supersonic modes propagating with velocities $3v_{\rm LR},5v_{\rm LR},\ldots$ for unequal-time correlations and with velocities $4v_{\rm LR},6v_{\rm LR},\ldots$ for equal-time correlations. These modes clearly violate the bounds~(\ref{lrbound}) and (\ref{lrbound2}) since the supersonic modes will eventually protrude into the tail beyond the light cone allowed by the bounds (characterized by $\xi$ and $\chi$). We note that the propagation of correlations in the unconditional dynamics $\hat{\rho}(t)=\sum_{n}P_{n}(t)\hat{\rho}^{(n)}(t)$ is still limited by the LR velocity (see the left-most panel in Fig.~1(c) in the main text), as it is consistent with the bounds. Here, the LR bound manifests itself as an exponential suppression of the supersonic modes due to  the exponentially decaying probability factor $P_{n}(t)$ which multiplies the full-counting dynamics $\hat{\rho}^{(n)}(t)$. To see this, in Fig.~\ref{figs3} we plot  (a) a typical profile of the correlation function in the full-counting dynamics and (b) the values of the supersonic contribution for different numbers of quantum jumps $n$. The latter shows that the supersonic contribution dwindles very rapidly (faster than exponential decrease) as $n$ increases.  Thus, the major contributions of the supersonic modes come from the trajectories with relatively small number $n$ of jumps (i.e., atomic loss). Meanwhile, the occurrence probability of such trajectories will  eventually be suppressed exponentially as a function of time $t$, where a substantial number of atoms are typically lost (see also the section ``Discussions" in the main text). It is this exponential suppression that recovers the LR bound in the overall unconditional density matrix $\hat{\rho}(t)=\sum_{n}P_{n}(t)\hat{\rho}^{(n)}(t)$.

\subsection{Derivation of the time-evolution equation of the exactly  solvable model}
\begin{figure}[t]
\includegraphics[width=180mm]{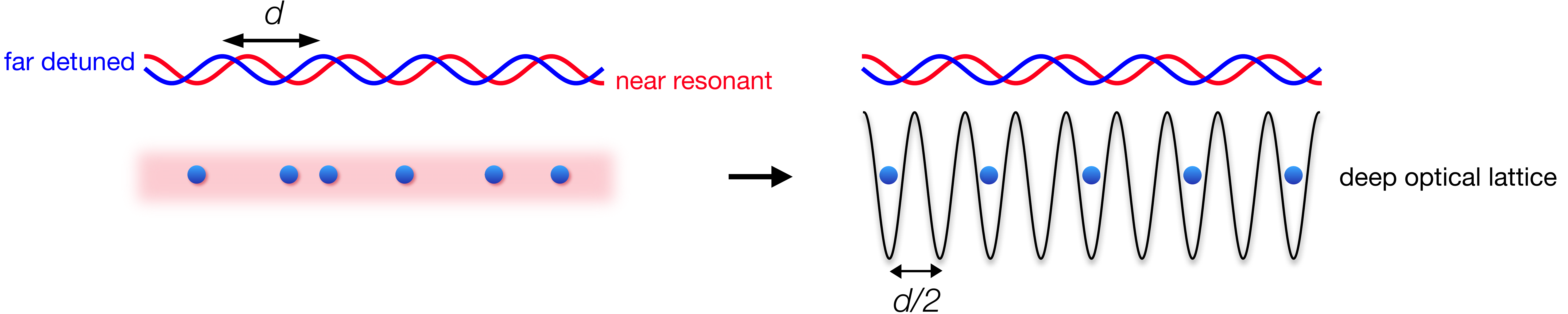}
\caption{\label{figs1}
\noindent{Schematic figures illustrating the model considered.} (left panel) The continuum model of one-dimensional ultracold atoms subject to two shallow optical lattices, one being far detuned (blue) and the other being near-resonant (red) to an atomic resonance. (right panel) Superimposing a deep optical potential (black) having the half periodicity $d/2$ and employing the tight-binding approximation, we obtain the lattice model considered in the main text.}
\end{figure}

Here we derive the time-evolution equation of the exactly solvable model introduced in the main text. We start from the continuum model of one-dimensional spinless fermionic atoms subject to two weak standing waves  with wavelength $\lambda$. One standing wave is far detuned from the atomic resonance and thus creates a shallow real potential $h_{0}\cos(2\pi x/d)$, where $h_{0}$ is the potential depth and $d=\lambda/2$ is the lattice spacing. The other is near resonant to the atomic resonance and creates a weak dissipative potential that leads to a one-body loss. These beams are superimposed and displaced from each other by $d/4$ (see Fig.~\ref{figs1}). Then, after adiabatically eliminating the dynamics of excited states \cite{MKO99,YA17nc}, the time evolution of ground-state atoms can be described by the following Lindblad master equation:

\eqn{\label{mastercS}
\frac{{d}\hat{\rho}}{{d}t}=-i\left(\hat{{\cal H}}_{{\rm eff}}\hat{\rho}-\hat{\rho}\hat{{\cal H}}_{{\rm eff}}^{\dagger}\right)+2\gamma_{0}\int dx\,\left[1+\sin\left(\frac{2\pi x}{d}\right)\right]\hat{\Psi}(x)\hat{\rho}\,\hat{\Psi}^{\dagger}(x),}
where
\eqn{
\hat{{\cal H}}_{{\rm eff}}\equiv\int dx\,\hat{\Psi}^{\dagger}(x)\left(-\frac{\nabla^{2}}{2m}+V_{\rm eff}(x)-i\gamma_{0}\right)\hat{\Psi}(x)\label{nonherS}
}
is an effective non-Hermitian Hamiltonian, $\hat{\Psi}(x)$ denotes the field operator of the atoms, $\gamma_{0}$ characterizes the strength of the dissipation that is determined by the intensity of the near-resonant light, and $V_{\rm eff}(x)=h_{0}\cos(2\pi x/d)-i\gamma_{0}\sin(2\pi x/d)$ is a complex effective potential \cite{AT03}. As in the main text, we set $\hbar=1$. 

We then superimpose a deep lattice potential with half periodicity $d/2$ (see Fig.~\ref{figs1}). Employing the standard procedure of the tight-binding approximation for the atomic field \cite{BI08}, we obtain the following master equation:

\eqn{
\frac{d\hat{\rho}(t)}{dt}=-i\left(\hat{H}_{\rm eff}\hat{\rho}-\hat{\rho}\hat{H}_{\rm eff}^{\dagger}\right)+{\cal J}[\hat{\rho}],\label{masterS}}
where
\eqn{
\hat{H}_{\mathrm{eff}}=-\sum_{l}\left[\left(J+(-1)^{l}i\gamma\right)\!\!\left(\hat{c}_{l+1}^{\dagger}\hat{c}_{l}+\hat{c}_{l}^{\dagger}\hat{c}_{l+1}\right)+(-1)^{l}h\hat{c}_{l}^{\dagger}\hat{c}_{l}\right]-2i\gamma w\hat{N}
}
is a tight-binding version of the effective Hamiltonian, and
\eqn{
\mathcal{J}[\hat{\rho}]=2\gamma\sum_{l}[2w\hat{c}_{l}\hat{\rho}\hat{c}_{l}^{\dagger}+(-1)^{l}(\hat{c}_{l}\hat{\rho}\hat{c}_{l+1}^{\dagger}+\hat{c}_{l+1}\hat{\rho}\hat{c}_{l}^{\dagger})]\label{jumpS}
}
is a super-operator acting on the density matrix $\hat{\rho}$, which describes a quantum jump process; $\hat{c}_{l}$ ($\hat{c}^{\dagger}_{l}$) is the annhilation (creation) operator of the atom at the site $l$, $J$ ($\gamma$) is the real (imaginary) hopping parameter, $h$ is the staggered on-site potential, $w$ is a factor determined by the depth of the deep lattice, and $\hat{N}=\sum_{l}\hat{c}_{l}^{\dagger}\hat{c}_{l}$ is the total atom-number operator. To ensure that the dynamical map (\ref{masterS}) is completely positive trace-preserving (CPTP) (i.e., the dynamics is Markovian), we must impose the condition $w\geq 1$. For the sake of concreteness,  we  assume $w=1$ below though the specific choice of its value is irrelevant to our findings discussed in the main text. 

\subsection{ Explicit expression of the exact solution of the lattice model}

We here provide technical details on the derivation for the solution of the full-counting dynamics discussed in the main text. To begin with, we diagonalize the effective Hamiltonian $\hat{H}_{\rm PT}$ (see Eq.~(4) in the main text). We divide the periodic chain of even length $L$ into two sublattices ($\hat{a}_{l}=\hat{c}_{2l}$ and $\hat{b}_{l}=\hat{c}_{2l+1}$ with $l=0,1,\ldots,L/2-1$) and introduce their Fourier transforms by

\begin{equation}
\hat{a}_{l}=\sqrt{\frac{2}{L}}\sum_{0\leq k<2\pi}\hat{a}_k e^{ikl},\;\;\;\hat{b}_{l}=\sqrt{\frac{2}{L}}\sum_{0\leq k<2\pi}\hat{b}_k e^{ikl},\;\;\;k=\frac{2\pi n}{(L/2)}\;\left(n=0,1,\ldots,\frac{L}{2}-1\right).
\end{equation}
Using these operators, we can rewrite $\hat{H}_{\rm PT}$ as follows:

\eqn{
\hat{H}_{\rm PT}&=&-\sum_{l=0}^{L-1}\!\left[\left(J\!+\!(-1)^{l}i\gamma\right)\!\!\left(\hat{c}_{l+1}^{\dagger}\hat{c}_{l}\!+\!\hat{c}_{l}^{\dagger}\hat{c}_{l+1}\right)\!+\!(-1)^{l}h\hat{c}_{l}^{\dagger}\hat{c}_{l}\right]\\
&=&\!
\sum_{0\leq k<2\pi}\!\left(\begin{array}{cc}
\hat{a}^{\dagger}_k & \hat{b}^{\dagger}_k \end{array}\right)\!\left(\begin{array}{cc}
-h & -J-i\gamma+(-J+i\gamma)e^{-ik}\\
-J-i\gamma+(-J+i\gamma)e^{ik} & h
\end{array}\right)\!\left(\begin{array}{c}
\hat{a}_k\\
\hat{b}_k
\end{array}\right).\label{pt2S}
}
Diagonalizing the $2\times 2$ matrix for each mode $k$, we obtain

\eqn{
\hat{H}_{\rm PT}=\sum_{k}\sum_{\lambda=\pm}\epsilon_{\lambda}(k)\hat{g}_{\lambda k}^{\dagger}\hat{f}_{\lambda k},\;\epsilon_{\pm}(k)=\pm\sqrt{h^{2}-4\gamma^{2}+2J'^2(1+\cos k)},\;\hat{g}_{\lambda k}^{\dagger}=\alpha_{\lambda}^{\rm R}(k)\hat{a}^{\dagger}_{k}+\beta_{\lambda}^{\rm R}(k)\hat{b}^{\dagger}_{k},\label{diagS}
}
where $\epsilon_{\lambda}(k)$ ($\lambda=\pm$) are two eigenvalues for each mode $k$ with $J'=\sqrt{J^2+\gamma^2}$, $\hat{g}_{\lambda k}^{\dagger}$ creates a right eigenvector of $\hat{H}_{\rm PT}$, i.e., $\hat{H}_{\mathrm{PT}}\hat{g}^{\dagger}_{\lambda k}|0\rangle=\epsilon_{\lambda}(k)\hat{g}^{\dagger}_{\lambda k}|0\rangle$, and ($\alpha_{\lambda}^{\rm R}(k), \beta_{\lambda}^{\rm R}(k))^{\rm T}$ are components of the corresponding right eigenvector of the $2\times 2$ non-Hermitian matrix in Eq.~(\ref{pt2S}). The operator $\hat{f}_{\lambda k}$ creates a left eigenvector of $\hat{H}_{\rm PT}$, i.e., $\langle 0|\hat{f}_{\lambda k}\hat{H}_{\mathrm{PT}}=\langle 0|\hat{f}_{\lambda k} \epsilon_{\lambda}(k)$, and its form is uniquely determined when we impose a generalized anticommutation relation $\{\hat{f}_{\lambda k},\hat{g}_{\lambda' k'}^{\dagger}\}=\delta_{k,k'}\delta_{\lambda,\lambda'}$.

We can similarly diagonalize the jump term (\ref{jumpS}), obtaining

\eqn{
{\cal J}[\hat{\rho}]=4\sum_{k}\sum_{\lambda=\pm}\gamma_{\lambda}(k)\hat{d}_{\lambda k}\,\hat{\rho}\,\hat{d}_{\lambda k}^{\dagger},\;\;\;\gamma_{\pm}(k)=\gamma\left(1\pm\left|\sin\left(\frac{k}{2}\right)\right|\right),
}
where we introduce the operators $\hat{d}_{+,k}=(-ie^{\frac{ik}{2}}\hat{a}_{k}+\hat{b}_{k})/\sqrt{2}$ and $\hat{d}_{-,k}=(ie^{\frac{ik}{2}}\hat{a}_{k}+\hat{b}_{k})/\sqrt{2}$.

We next derive the solution of the full-counting dynamics discussed in the main text. To do so, we introduce the interaction picture by

\begin{equation}
\hat{\tilde{\rho}}(t)=e^{i\hat{H}_{{\rm eff}}t}\hat{\rho}(t)e^{-i\hat{H}_{{\rm eff}}^{\dagger}t},\;\;\;\hat{\tilde{d}}_{\lambda k}(t)=e^{i\hat{H}_{{\rm eff}}t}\hat{d}_{\lambda k}e^{-i\hat{H}_{{\rm eff}}t}.
\end{equation}
Then, the time-evolution equation becomes

\begin{equation}
\frac{d\hat{\tilde{\rho}}(t)}{dt}=4\sum_{k}\sum_{\lambda=\pm}\gamma_{\lambda}(k)\hat{\tilde{d}}_{\lambda k}(t)\hat{\tilde{\rho}}(t)\hat{\tilde{d}}_{\lambda k}^{\dagger}(t).\label{drdS}
\end{equation}
For the sake of concreteness, we assume that $N=L/2$ particles are present at time $t=0$ as considered in the main text. Introducing a projector $\hat{\cal P}_{n}$ onto the subspace containing $N-n$ particles, we denote an unnormalized density matrix accompanying $n$ quantum jumps by $\hat{\varrho}^{(n)}(t)=\hat{\cal P}_{n}\hat{\rho}(t)\hat{\cal P}_{n}$. Integrating out Eq.~(\ref{drdS}) with the initial condition $\hat{\varrho}^{(n)}(0)=0$ for $n<N$ and noting the relation $[\hat{\cal P}_{n},\hat{H}_{\rm eff}]=0$, we obtain the following recursion relation:

\begin{equation}
\hat{\tilde{\varrho}}^{(n)}(t)=4\int_{0}^{t}d\tau\sum_{k,\lambda}\gamma_{\lambda}(k)\hat{\tilde{d}}_{\lambda k}(\tau)\hat{\tilde{\varrho}}^{(n-1)}(\tau)\hat{\tilde{d}}_{\lambda k}^{\dagger}(\tau).\label{eq:1}
\end{equation}
Solving the recursion relation (\ref{eq:1}) iteratively, we obtain the formal solution of $\hat{\tilde{\varrho}}^{(n)}(t)$ as
\eqn{
\hat{\tilde{\varrho}}^{(n)}(t) &=&4^{n}\int_{0}^{t}dt_{n}\cdots\int_{0}^{t_{2}}dt_{1}\nonumber\\
&&\times\sum_{k_{1}\lambda_{1},\ldots,k_{n}\lambda_{n}}\gamma_{\lambda_{1}}(k_{1})\cdots\gamma_{\lambda_{n}}(k_{n})\hat{\tilde{d}}_{\lambda_{n}k_n}(t_{n})\cdots\hat{\tilde{d}}_{\lambda_{1}k_1}(t_{1})\hat{\tilde{\varrho}}^{(0)}(t_1)\hat{\tilde{d}}_{\lambda_{1}k_1}^{\dagger}(t_{1})\cdots\hat{\tilde{d}}_{\lambda_{n}k_n}^{\dagger}(t_{n})\nonumber\\
&=&\frac{4^{n}}{n!}\int_{0}^{t}dt_{n}\cdots\int_{0}^{t}dt_{1} \nonumber\\
&&\times\!\sum_{k_{1}\lambda_{1},\ldots,k_{n}\lambda_{n}}\gamma_{\lambda_{1}}(k_{1})\cdots\gamma_{\lambda_{n}}(k_{n})\overrightarrow{T}\left[\hat{\tilde{d}}_{\lambda_{n}k_n}(t_{n})\cdots\hat{\tilde{d}}_{\lambda_{1}k_1}(t_{1})\right]\!\hat{\rho}(0)\!\overleftarrow{T}\left[\hat{\tilde{d}}_{\lambda_{1}k_1}^{\dagger}(t_{1})\cdots\hat{\tilde{d}}_{\lambda_{n}k_n}^{\dagger}(t_{n})\right]\!,\label{formalS}
}
where we use $\hat{\tilde{\varrho}}^{(0)}(t_{1})=\hat{\rho}(0)$ in obtaining the second equality, and $\overrightarrow{T}$ ($\overleftarrow{T}$) denotes the time-ordering (anti-time-ordering) operator. To perform the time integration, let us simplify the following time-dependent operator $\hat{\tilde{d}}_{\lambda k}(t)$:

\begin{equation}
\hat{\tilde{d}}_{\lambda k}(t)=e^{i\hat{H}_{{\rm eff}}t}\hat{d}_{\lambda k}e^{-i\hat{H}_{{\rm eff}}t}=e^{-2\gamma t}e^{i\hat{H}_{\rm PT}t}\hat{d}_{\lambda k}e^{-i\hat{H}_{\rm PT}t}\label{eq:deq}.
\end{equation}
Because $\hat{H}_{\rm PT}$ is quadratic in fermionic operators (see Eq.~(\ref{pt2S})), we can solve Eq.~(\ref{eq:deq}) by introducing an eigenoperator $\hat{\Lambda}_{\eta k}$ satisfying the relation $[\hat{\Lambda}_{\eta k},\hat{H}_{\rm PT}]=\epsilon_{\eta}(k)\hat{\Lambda}_{\eta k}\;\;(\eta=\pm)$. We thus obtain

\begin{equation}
\hat{\tilde{d}}_{\lambda k}(t)=e^{-2\gamma t}\sum_{\eta=\pm}c_{\eta\lambda}\hat{\Lambda}_{\eta k}e^{-i\epsilon_{\eta}(k)t},
\end{equation}
where $c_{\eta\lambda}$ are the expansion coefficients of $\hat{d}_{\lambda k}$ with respect to $\hat{\Lambda}_{\eta k}$'s. Using a right eigenvector $(\alpha_{\lambda}^{\rm R}(k),\beta_{\lambda}^{\rm R}(k))^{\rm T}$ of $\hat{H}_{\rm PT}$ (see Eq.~(\ref{diagS})), an explicit expression of the eigenoperators can be given as $\hat{\Lambda}_{\eta k}=\alpha_{\eta}^{\rm R}(-k)\hat{a}_{k}+\beta_{\eta}^{\rm R}(-k)\hat{b}_{k}$. We then consider the
following time integration:

\begin{align}
\int_{0}^{t}d\tau\cdots\hat{\tilde{d}}_{\lambda k}(\tau)\cdots\hat{\tilde{d}}_{\lambda k}^{\dagger}(\tau)\cdots & =\int_{0}^{t}d\tau\sum_{\eta\eta'}c_{\eta\lambda}c_{\eta'\lambda}^{*}e^{-4\gamma\tau-i\epsilon_{\eta}(k)\tau+i\epsilon_{\eta'}(k)\tau}\cdots\hat{\Lambda}_{\eta k}\cdots\hat{\Lambda}_{\eta' k}^{\dagger}\cdots\nonumber\\
 & =\sum_{\eta\eta'}c_{\eta\lambda}c_{\eta'\lambda}^{*}\frac{1-e^{-4\gamma t-i\epsilon_{\eta}(k)t+i\epsilon_{\eta'}(k)t}}{4\gamma+i(\epsilon_{\eta}(k)-i\epsilon_{\eta'}(k))}\cdots\hat{\Lambda}_{\eta k}\cdots\hat{\Lambda}_{\eta' k}^{\dagger}\cdots.
\end{align}
From Eq.~(\ref{formalS}), and by introducing the $2\times 2$ matrix $\gamma_{\eta\eta'}^{c}(k)=\sum_{\lambda}\gamma_{\lambda}(k)c_{\eta\lambda}c_{\eta'\lambda}^{*}$, we obtain

\eqn{
\hat{\tilde{\varrho}}^{(n)}(t)=\sum_{\eta_{1}\eta'_{1}k_{1}\cdots\eta_{n}\eta'_{n}k_{n}}\frac{1}{n!}\left[\prod_{i=1}^{n}\gamma_{\eta_{i}\eta'_{i}}^{c}(k_{i})\frac{1-e^{-4\gamma t-i\epsilon_{\eta_{i}}(k_{i})t+i\epsilon_{\eta'_{i}}(k_{i})t}}{\gamma+i(\epsilon_{\eta_{i}}(k_{i})-\epsilon_{\eta'_{i}}(k_{i}))/4}\right]\hat{\Lambda}_{\eta_{n} k_{n}}\cdots\hat{\Lambda}_{\eta_{1} k_1}\hat{\rho}(0)\hat{\Lambda}_{\eta'_{1} k_1}^{\dagger}\cdots\hat{\Lambda}_{\eta'_{n} k_n}^{\dagger}.\nonumber\\
}
Transforming back to the Schr\"odinger picture by using 
\begin{equation}
\hat{\varrho}^{(n)}(t)=e^{-i\hat{H}_{{\rm eff}}t}\hat{\tilde{\varrho}}^{(n)}(t)e^{i\hat{H}_{{\rm eff}}^{\dagger}t}=e^{-4\gamma(N-n)t}e^{-i\hat{H}_{\rm PT}t}\hat{\varrho}^{(n)}(t)e^{i\hat{H}_{\rm PT}^{\dagger}t}
\end{equation}
and the relation $e^{-i\hat{H}_{\rm PT}t}\hat{\Lambda}_{\eta k}e^{i\hat{H}_{\rm PT}t}=\hat{\Lambda}_{\eta k}e^{i\epsilon_{\eta}(k)t}$, we obtain the solution of the full-counting dynamics: 

\eqn{
\hat{\varrho}^{(n)}(t)\!=\!\!\sum_{\eta_{1}\eta'_{1}k_{1}\cdots\eta_{n}\eta'_{n}k_{n}}\!\frac{e^{-4\gamma(N-n)t}}{n!}\!\left[\prod_{i=1}^{n}{\cal D}_{\eta_{i}\eta'_{i}}(k_{i};t) \right]\!\!\hat{\Lambda}_{\eta_{n}k_n}\!\cdots\!\hat{\Lambda}_{\eta_{1}k_1}e^{-i\hat{H}_{\rm PT}t}\hat{\rho}(0)e^{i\hat{H}_{\rm PT}^{\dagger}t}\hat{\Lambda}_{\eta'_{1}k_1}^{\dagger}\!\cdots\!\hat{\Lambda}_{\eta'_{n}k_n}^{\dagger},\label{solS}
}
where we introduce the $2\times 2$ Hermitian matrix ${\cal D}_{\eta\eta'}$ by
\eqn{
{\cal{D}}_{\eta\eta'}(k;t)=\gamma_{\eta\eta'}^{c}(k)\frac{e^{i\epsilon_{\eta}(k)t-i\epsilon_{\eta'}(k)t}-e^{-4\gamma t}}{\gamma+i(\epsilon_{\eta}(k)-\epsilon_{\eta'}(k))/4}.
}

In practice, to calculate the nonequilibrium properties of the system such as correlation functions, we proceed as follows. First, we diagonalize the operators $\hat{\Lambda}_{\eta k}$ and $\hat{\Lambda}^{\dagger}_{\eta' k}$ in Eq.~(\ref{solS}) with respect to the indices $\eta$ and $\eta'$. To this end, for each time $t$ and wavevector $k$, we numerically diagonalize the following $2\times2$ Hermitian matrix:

\begin{equation}
\left(\begin{array}{cc}
\sum_{\eta\eta'}{\cal D}_{\eta\eta'}(k;t)\alpha_{\eta}^{\rm R}(-k)\alpha_{\eta'}^{*{\rm R}}(-k) & \sum_{\eta\eta'}{\cal D}_{\eta\eta'}(k;t)\beta_{\eta}^{\rm R}(-k)\alpha_{\eta'}^{*{\rm R}}(-k)\\
\sum_{\eta\eta'}{\cal D}_{\eta\eta'}(k;t)\alpha_{\eta}^{\rm R}(-k)\beta_{\eta'}^{*{\rm R}}(-k) & \sum_{\eta\eta'}{\cal D}_{\eta\eta'}(k;t)\beta_{\eta}^{\rm R}(-k)\beta_{\eta'}^{*{\rm R}}(-k)
\end{array}\right).\label{eq:mat}
\end{equation}
Using its two real eigenvalues $\lambda_{\pm,k}(t)$ and the corresponding orthonormal eigenvectors $\hat{v}_{\pm,k}(t)$, we can simplify Eq.~(\ref{solS}) as follows:

\eqn{
\hat{\varrho}^{(n)}(t)\!=\!\!\frac{e^{-4\gamma(N-n)t}}{n!}\!\!\sum_{\eta_{1}k_{1}\cdots\eta_{n}k_{n}}\!\left[\prod_{i=1}^{n}\lambda_{\eta_{i}k_{i}}(t)\right]\!\hat{v}_{\eta_{n}k_n}(t)\!\cdots\!\hat{v}_{\eta_{1}k_1}(t)e^{-i\hat{H}_{\rm PT}t}\hat{\rho}(0)e^{i\hat{H}_{\rm PT}^{\dagger}t}\hat{v}_{\eta_{1}k_1}^{\dagger}(t)\!\cdots\!\hat{v}_{\eta_{n}k_n}^{\dagger}(t). \label{simpleS}
}
The time evolution $e^{-i\hat{H}_{PT}t}\hat{\rho}(0)e^{i\hat{H}_{PT}^{\dagger}t}$ can be calculated by using Eq.~(\ref{diagS}). Denoting the initial state as $\hat{\rho}(0)=|\Psi_{0}\rangle\langle\Psi_{0}|$ and expanding it in terms of the right eigenvectors $|\Psi_{0}\rangle=\prod_{k}[\sum_{\lambda}\psi_{\lambda k}\hat{g}_{\lambda k}^{\dagger}]|0\rangle$,  the time evolution is given by
\begin{equation}\label{PTS}
|\Psi_{t}\rangle=e^{-i\hat{H}_{\rm PT}t}|\Psi_{0}\rangle=\prod_{k}\left[\sum_{\lambda}\psi_{\lambda k}e^{-i\epsilon_{\lambda}(k)t}\hat{g}_{\lambda k}^{\dagger}\right]|0\rangle=\prod_{k}\left[\sum_{\eta}\psi_{\eta}^{v}(k;t)\hat{v}_{\eta k}^{\dagger}(t)\right]|0\rangle.
\end{equation}
In obtaining the last equality, we have expanded the time-dependent state in the basis of $\hat{v}_{\pm,k}(t)$ and introduced the corresponding expansion coefficients $\psi_{\eta}^{v}(k;t)$. Combining Eqs.~(\ref{simpleS}) and (\ref{PTS}), we can obtain the following  expression for the trace of an unnormalized density matrix:

\begin{equation}
{\rm Tr}\left[\hat{\varrho}^{(n)}(t)\right]=e^{-4\gamma(N-n)t}\left[\prod_{k}{\cal N}_{k}(t)\right]\sigma_{n}\left(\{f^{v}_{k}(t)\}\right),
\end{equation}
where we introduce the time-dependent norm factor ${\cal N}_{k}(t)=\sum_{\eta}|\psi_{\eta k}^{v}(t)|^{2}$ of each mode $k$, and  $\sigma_{n}$ denotes the $n$ th symmetric polynomial of $f^{v}_{k}(t)\equiv\sum_{\eta}\lambda_{\eta k}(t)|\psi_{\eta k}^{v}(t)|^{2}/{\cal N}_{k}(t)$:

\begin{equation}
\sigma_{n}\left(\{f^{v}_{k}(t)\}\right)=\frac{(-1)^{n}}{(N-n)!}\frac{d^{N-n}}{dx^{N-n}}\Biggr|_{x=0}\prod_{k}\left(x-f^{v}_{k}(t)\right).\label{eq:mathe}
\end{equation}
We note that $\{f_{k}^{v}(t)\}$ forms a set of $N$ variables with  $k=0,2\pi/N,\ldots,2\pi(N-1)/N$.
%While the symmetric polynomial $\sigma_{n}$ in general contains $\left(\begin{array}{c} N\\n\end{array}\right)$ different terms, the relation (\ref{eq:mathe}) allows its efficient numerical evaluation. 

The full-counting equal-time correlation is now given as

\eqn{
C^{(n)}(l,t)&=&\frac{{\rm Tr}\left[\hat{c}_{l}^{\dagger}\hat{c}_{0}\hat{\varrho}^{(N-n)}(t)\right]}{{\rm Tr}\left[\hat{\varrho}^{(N-n)}(t)\right]}\nonumber\\
&=&\frac{2}{L}\sum_{k}e^{-ik\lceil l/2\rceil}\left[\frac{\sum_{\lambda}\psi_{\lambda k}^{*}(t)O_{\lambda}^{*{\rm R}}(k)\times\sum_{\lambda}\psi_{\lambda k}(t)\alpha_{\lambda}^{\rm R}(k)}{{\cal N}_{k}(t)}\right]\frac{\sigma_{n}\left(\{f^{v}_{k'}(t)\}_{k'\neq k}\right)}{\sigma_{n}\left(\{f^{v}_{k'}(t)\}\right)},
}
where we choose $O=\alpha\;(\beta)$ when $l$
is even (odd) and introduce $\psi_{\lambda k}(t)=\psi_{\lambda k}e^{-i\epsilon_{\lambda}(k)t}$ and
\begin{equation}
\sigma_{n}\left(\{f^{v}_{k'}(t)\}_{k'\neq k}\right)=\frac{(-1)^{n}}{(N-1-n)!}\frac{d^{N-1-n}}{dx^{N-1-n}}\Biggr|_{x=0}\prod_{k'\neq k}\left(x-f^{v}_{k'}(t)\right).
\end{equation}

Finally, for the null-jump case $n=0$, we can  further simplify the expressions of correlation functions. For example, the unequal-time correlation defined by $\tilde{C}^{(0)}(l,t)=\langle\Psi_{0}|\hat{c}_{l}^{\dagger}(t)\hat{c}_{0}(0)|\Psi_{0}\rangle/\langle\Psi_{t}|\Psi_{t}\rangle$ with $\hat{c}^{\dagger}_{l}(t)\equiv e^{i\hat{H}_{\mathrm{PT}}^{\dagger}t}\hat{c}_{l}^{\dagger}e^{-i\hat{H}_{\mathrm{PT}}t}$ can be expressed as
\eqn{
\tilde{C}^{(0)}(l,t)=\frac{2}{L}\sum_{k}\sum_{\lambda=\pm}O_{\lambda k}\frac{\psi_{\lambda k}^{*}\,e^{i\epsilon_{\lambda}(k)t-ik\lceil l/2\rceil}}{{\cal N}_{k}(t)}\;\;\;\;{\rm with}\;\;\;\;\;O_{\lambda k}=O_{\lambda}^{*{\rm R}}(k)\sum_{\eta}\psi_{\eta k}\alpha_{\eta}^{\rm R}(k),
}
which gives Eq.~(5) in the main text. 

\subsection{Chiral structure due to coalescence of eigenvectors near the exceptional point}
\begin{figure}[t]
\includegraphics[width=120mm]{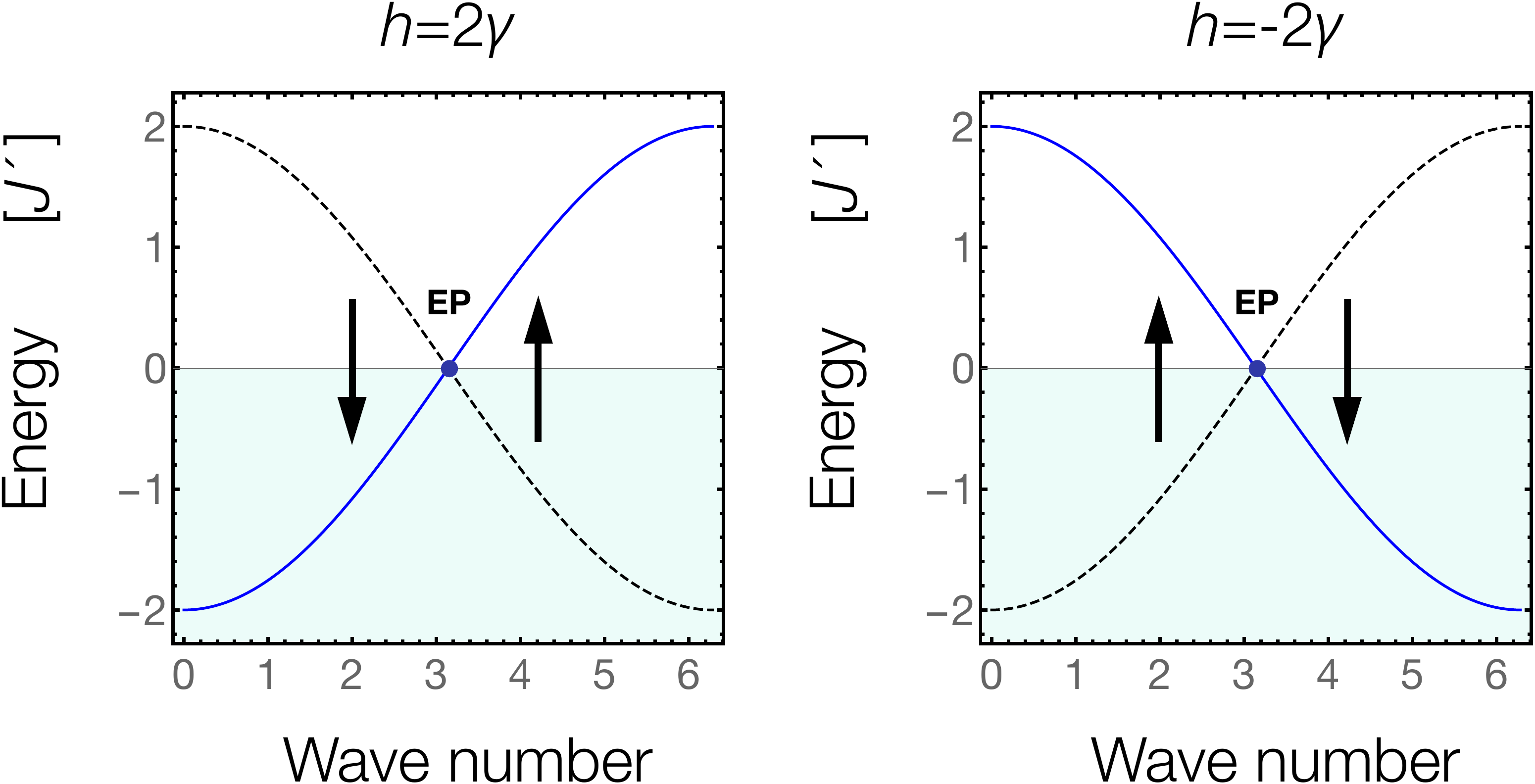}
\caption{Coalescence of eigenvectors near the exceptional point. (left panel) When the postquench Hamiltonian is near the spectral transition point, the gapless point at $k=\pi$ forms an exceptional point (EP). In the vicinity of EP, the two eigenvectors in different bands coalesce into the one associated with the band dispersion having positive group velocities (blue solid curve). Since the lower band is initially populated (shaded region), this coalescence leads to imbalanced effective population in the band having positive group velocities in $k<\pi$, resulting in the pronounced propagation of correlations in the positive direction as found in the main text. (right panel) In contrast, if the gain-loss structure is reversed (if we set $h=-2\gamma$), the band having negative group velocities (blue solid curve) is dominantly populated in $k>\pi$. This results in the  pronounced propagation in the negative direction.\label{figs2}}
\end{figure}

We here explain in detail how the chiral structure is induced by the coalescence of two eigenvectors of different bands, and provide simple analytical expressions of the eigenvectors in the vicinity of the exceptional point. Let us start from the Hermitian case, i.e., the Hamiltonian with $\gamma=h=0$ in Eq.~(\ref{pt2S}). Without loss of generality, we set $J=1$ throughout this section. There are two band dispersions: one has a positive group velocity ($\epsilon_{>}(k)=2\sin(k/2)$) and the other has a negative group velocity ($\epsilon_{<}(k)=-2\sin(k/2)$). The corresponding eigenvectors are given by diagonalizing the 2$\times$2 Hermitian matrices in Eq.~(\ref{pt2S}) with $\gamma=h=0$ and $J=1$. In the vicinity of the gapless point at $k=\pi$, we obtain the results
\eqn{
\bm{c}_{>}(\delta k)&=&\frac{1}{\sqrt{2}}\left(\begin{array}{c}
-i-\frac{\delta k}{2}\\
1
\end{array}\right)+O\left((\delta k)^2\right),\label{clposi}\\
\bm{c}_{<}(\delta k)&=&\frac{1}{\sqrt{2}}\left(\begin{array}{c}
i+\frac{\delta k}{2}\\
1
\end{array}\right)+O\left((\delta k)^2\right),\label{clnega}
}
where $\delta k=k-\pi$ is the displacement satisfying $|\delta k|\ll 1$ and $\bm{c}_{>(<)}$ is the eigenvector of the band dispersion having the positive (negative) group velocity. Since the lower band ($\epsilon(k)<0$) is filled in the initial ground state, the eigenvector $\bm{c}_{>}(\delta k)$ $\left(\bm{c}_{<}(\delta k)\right)$ is populated for $\delta k<0$ $\left(\delta k>0\right)$ at the initial time (see shaded region in Fig.~\ref{figs2}). 

We next consider the postquench Hamiltonian, i.e., the non-Hermitian Hamiltonian in Eq.~(\ref{pt2S}) with nonzero $\gamma$ satisfying $h=2\gamma$. In this case, two eigenvectors of different bands coalesce at the exceptional point with $k=\pi$. In the vicinity of the exceptional point, we obtain simple expressions of right eigenvectors of the 2$\times$2 non-Hermitian matrices in Eq.~(\ref{pt2S}) with $h=2\gamma$ and $J=1$ as follows:

\eqn{
\bm{c}^{\rm R}_{>}(\delta k)&=&\frac{1}{\sqrt{2}}\left(\begin{array}{c}
-i\\
1
\end{array}\right)+\frac{1}{4\sqrt{2}\gamma}\left(\begin{array}{c}
-i(1-2i\gamma-\sqrt{1+\gamma^2})\\
\sqrt{1+\gamma^2}-1
\end{array}\right)\delta k+O\left((\delta k)^2\right),
\\
\bm{c}^{\rm R}_{<}(\delta k)&=&\frac{1}{\sqrt{2}}\left(\begin{array}{c}
-i\\
1
\end{array}\right)+\frac{1}{4\sqrt{2}\gamma}\left(\begin{array}{c}
-i(1-2i\gamma+\sqrt{1+\gamma^2})\\
-\sqrt{1+\gamma^2}-1
\end{array}\right)\delta k+O\left((\delta k)^2\right),
}
which are valid for $|\delta k|\ll \min(\gamma,1)$. Here $\bm{c}^{\rm R}_{>(<)}$ is the right eigenvector of the band dispersion having the positive (negative) group velocity. Note that in the limit of $\delta k\to 0$ these two eigenvectors coalesce into one having a positive group velocity given in Eq.~(\ref{clposi}) (see the left panel in Fig.~\ref{figs2}). This coalescence of eigenvectors near the exceptional point leads to the imbalanced effective populations of quasiparticles having positive group velocities in $k<\pi$ (see Fig.~2(d) in the main text), resulting in the pronounced propagation of correlations in the positive direction as discussed in the main text. We remark that if the gain-loss structure is reversed, i.e., if we set $h=-2\gamma$, the two right eigenvectors can be shown to coalesce into one having a negative group velocity given in Eq.~(\ref{clnega}) (see the right panel in Fig.~\ref{figs2}), leading to the pronounced propagation in the negative direction. 

Finally, we explain why the discontinuity and the divergence of the effective band populations at $k=\pi$ (see Fig.~2(d) in the main text) are caused by  singularities in left eigenvectors at the exceptional point. To see this, let us discuss left eigenvectors of the $2\times 2$ non-Hermitian matrices in Eq.~(\ref{pt2S}) with $h=2\gamma$ near the exceptional point:

\eqn{
\bm{c}_{>}^{\rm L}(\delta k)&=&\!\frac{2\gamma}{\sqrt{2(1+\gamma^2)}}\left(\begin{array}{c}
i\\
1
\end{array}\right)\frac{1}{\delta k}\!+\!\frac{1}{2\sqrt{2(1+\gamma^2)}}\left(\begin{array}{c}
-i(1+2i\gamma+\sqrt{1+\gamma^2})\\
\sqrt{1+\gamma^2}+1
\end{array}\right)\!+\!O\left(\delta k\right),\label{opposi}\\
\bm{c}_{<}^{\rm L}(\delta k)\!\!&=&\!\!\!-\frac{2\gamma}{\sqrt{2(1+\gamma^2)}}\left(\begin{array}{c}
i\\
1
\end{array}\right)\!\frac{1}{\delta k}\!+\!\frac{1}{2\sqrt{2(1+\gamma^2)}}\left(\!\begin{array}{c}
-i(-1-2i\gamma+\sqrt{1+\gamma^2})\\
\sqrt{1+\gamma^2}-1
\end{array}\!\right)\!\!+\!O\!\left(\delta k\right),\label{opnega}
}
where $\bm{c}_{>(<)}^{\rm L}$ are left eigenvectors of the band dispersions having the positive (negative) group velocity. These expressions are valid for $|\delta k|\ll{\rm min}(\gamma,1)$. The divergence of the left eigenvectors in the limit $\delta k\to 0$ originates from the vanishing inner product between the right and left eigenvectors at the exceptional point \cite{HWD01,AR12}. To understand how this divergence leads to the discontinuity of the effective band populations $\tilde{n}_{\lambda k}$, we recall that $\tilde{n}_{\lambda k}$ is proportional to the square of the expansion coefficient $\psi_{\lambda k}$ of the initial ground state in terms of the right eigenvectors for the postquench non-Hermitian matrices, i.e., $n_{\lambda k}\propto |\psi_{\lambda k}|^2$ with $|\Psi_{0}\rangle=\prod_{k}[\sum_{\lambda}\psi_{\lambda k}\hat{g}_{\lambda k}^{\dagger}]|0\rangle$. We then obtain the following relations:

\eqn{
{\rm For\;\;}\delta k<0&:&\;\;\;\begin{cases}
\tilde{n}_{+,k}\propto\left|\bm{c}^{\dagger{\rm L}}_{<}(\delta k)\cdot\bm{c}_{>}(\delta k)\right|^2\sim 0
\\
\tilde{n}_{-,k}\propto\left|\bm{c}^{\dagger{\rm L}}_{>}(\delta k)\cdot\bm{c}_{>}(\delta k)\right|^2\sim 1,
\end{cases}\label{casen}
\\
{\rm For\;\;}\delta k>0&:&\;\;\;
\begin{cases}
\tilde{n}_{+,k}\propto\left|\bm{c}^{\dagger{\rm L}}_{>}(\delta k)\cdot\bm{c}_{<}(\delta k)\right|^2\simeq \frac{\gamma^2}{1+\gamma^2}\frac{1}{\delta k^2}
\\
\tilde{n}_{-,k}\propto\left|\bm{c}^{\dagger{\rm L}}_{<}(\delta k)\cdot\bm{c}_{<}(\delta k)\right|^2\simeq \frac{\gamma^2}{1+\gamma^2}\frac{1}{\delta k^2}.
\end{cases}\label{casep}
}
Here we note that the expansion coefficients $\psi_{\lambda k}$ in terms of right eigenvectors can be obtained by taking the inner product between the corresponding left eigenvectors and the initial ground state since the left and right eigenvectors satisfy the orthonormal condition $\{\hat{f}_{\lambda k},\hat{g}^{\dagger}_{\lambda',k'}\}=\delta_{k,k'}\delta_{\lambda,\lambda'}$. Equation~(\ref{casen}) shows that the populations $\tilde{n}_{\pm,k}$ remain finite if we approach  the exceptional point $k=\pi$ from below. This is because the diverging contribution $(i,1)^{\rm T}$ in the left eigenvectors in Eqs.~(\ref{opposi}) and (\ref{opnega}) is orthogonal to the leading contribution $(-i,1)^{\rm T}$ of $\bm{c}_{>}(\delta k)$ in Eq.~(\ref{clposi}). In contrast, if we approach the exceptional point from above, the populations $\tilde{n}_{\pm k}$ diverge in $\delta k\to 0$ as shown in Eq.~(\ref{casep}) since the diverging contribution $(i,1)^{\rm T}$ in Eqs.~(\ref{opposi}) and (\ref{opnega}) is parallel to the leading contribution of $\bm{c}_{<}(\delta k)$ in Eq.~(\ref{clnega}).

\end{document}